\documentclass[twocolumn,showpacs,aps,prl]{revtex4}  % preprint OR twocolumn
\usepackage{epsfig,amssymb}

\begin{document}

\title{Experimental studies of equilibrium vortex properties in a Bose-condensed gas}
\author{I. Coddington, P. C. Haljan, P. Engels, V. Schweikhard, S. Tung  and E.~A. Cornell$^\ast$}
\affiliation{JILA, National Institute of Standards and Technology and Department of Physics, \\
University of Colorado, Boulder, Colorado 80309-0440}
\date{\today}

\begin{abstract}
We characterize several equilibrium vortex effects in a rotating
Bose-Einstein condensate.  Specifically we attempt precision
measurements of vortex lattice spacing and the vortex core size
over a range of condensate densities and rotation rates. These
measurements are supplemented by numerical simulations, and both
experimental and numerical data are compared to theory. Finally,
we study the effect of the centrifugal weakening of the trapping
spring constants on the critical temperature for quantum
degeneracy and the effects of finite temperature on vortex
contrast.
\end{abstract}

\pacs{03.75.Lm, 67.90.+z, 67.40.Vs, 32.80.Pj} \maketitle

\section{I. Introduction}%\label{SecIntro}
%\subsection{A. Overview}%\label{overview}

After the initial observations of vortex lattices in Bose-Einstein
condensed gases \cite{DalibardLatt,KetterleLatt,Paul,FootVortex},
most of the experimental work has focused on dynamical behavior of
vortices and lattices, including Kelvons
\cite{DalibardKv,SandroKv,StoofKv}, Tkachenko waves
\cite{AnglinTk,JilaTk,BaymTk,BigelowTk,MachidaTk,BaymTk2}, and
various nonequilibrium effects \cite{Jilastripes,GiantVortex}.
Equilibrium properties, in contrast, have been relatively
neglected by experimenters. This imbalance is not indicative of a
lack of interesting physics in equilibrium behavior, but simply
reflects the usual experimentalist's preference for measuring
spectra rather than static structure. Theorists, on the other
hand, have investigated equilibrium properties extensively
\cite{BaymLLL,Dan,FederNv,FetterCores,FetterLLL,HoLLL,AllanQH,MachidaCore,SalomaaCore,BaymLatt,ReadLatt,SandroTc},
and our purpose in this paper is to partially redress this
imbalance with a series of experimental studies focusing on
equilibrium properties of rotating condensates.
\par
The vortex lattice in a rotating Bose-condensed gas naturally
organizes into a regular triangular lattice, or Abrikosov lattice,
originally observed in superconductors.  The lattice can be well
characterized by the nearest-neighbor lattice spacing and by the
radius of each vortex core ($b$ and $r_{v}$ respectively for the
purpose of this paper). The nearest-neighbor lattice spacing, $b$,
is generally thought to be determined only by the rotation rate
when in the high-rotation regime where the rotating BEC exhibits
nearly rigid-body behavior. Numerical work \cite{FederNv} and
early analytical work \cite{AnglinTk}, however, suggests that this
rigid-body assumption yields lattice constants that are smaller
than would be seen in the case of a finite-size trapped BEC.
Recent work by Sheehy and Radzihovsky \cite{Dan} has tackled
analytically this discrepancy and found it to be a necessary
consequence of the inhomogeneous density profile of the
condensate. With this theory they address the question of why the
lattice is so remarkably regular given the condensate density
profile. They also derive a small, position-dependent,
inhomogeneity-induced correction term to the lattice spacing. An
interesting implication of this theory is that the vortices must
move slightly faster than the surrounding superfluid even near the
rigid-body limit. More striking still is the prediction that the
superfluid should exhibit a radially-dependant angular velocity
(or radial shear flow), which directly follows from their
calculation of inhomogeneous vortex density. While a differential
rotation rate is not directly observable in our system, the
position-dependent variation of the nearest-neighbor lattice
spacing is studied in Sec.~II below. It should also be noted that
the inhomogeneity in the areal density of vortices, predicted in
Ref. \cite{Dan}, can also be derived in the limit of the lowest
Landau level (LLL). This property of the LLL was first brought to
our attention by A.H. MacDonald and has been the subject of two
recent publications by Watanabe \emph{et al.} \cite{BaymLatt} and
Cooper \emph{et al.} \cite{ReadLatt}.
\par
The second effect we study in this paper concerns the core size of
the vortices.  Once rotation rate and density are fixed, the
vortex core size is a length scale that the condensate chooses on
its own.  In this sense vortex core size constitutes a fundamental
property of the system and has therefore been the subject of much
theoretical work~\cite{FetterCores,MachidaCore,SalomaaCore}. By
analogy to superfluid $^{4}$He, the core size is dictated by the
atomic interactions and is of order of the healing length. For our
system the healing length is only one and a half times the average
interatomic spacing. Because of this diluteness one might wonder
if there are certain regimes of sufficiently low or high density
where one would see a deviation from mean-field theory.
Investigation of core size makes up Sec.~III of this paper.
\par
It has been predicted~\cite{HoLLL} that at higher rotation the
condensate begins to enter the LLL regime where the condensate
wave function is constrained to occupy only the LLL harmonic
oscillator states.  Our condensate is strikingly different from
quantum Hall systems also associated with the LLL in that the
mean-field approximation is still a valid way to account for
interactions. In this case the vortex core-size to lattice-spacing
ratio saturates at the LLL limit~\cite{Fischer,JilaLLL}. As Baym
and Pethick~\cite{BaymLLL} note this requires the cores to deviate
from the Thomas-Fermi prediction in shape and size.
\par
Employing a numerically generated Gross-Pitaevskii (GP) wave
function, we study this transition and monitor the core size and
shape over a range of conditions. These simulations show a smooth
transition from the Thomas-Fermi regime to the LLL regime.
Additionally we use numerical analysis to study the shape of the
vortex cores and verify that when the core-area saturation occurs
the vortex-core wave function is the one predicted for the LLL
regime. This result suggests that the fractional core area is a
possible means to probe the transition to the LLL regime. To test
this idea, we compare simulations with experimental data.  This
comparison and possible systematic errors in these experimental
measurements are discussed in Sec.~III~\small{B}\normalsize.
\par
In Sec.~IV we examine the rotational suppression of quantum
degeneracy. This effect is due to centrifugal forces weakening the
radial trap-spring constants. Weaker spring constants then lead,
in a straight forward way, to a lower critical temperature for
fixed numbers of atoms.
\par
Finally in Sec.~V we examine a proposal that a measurement of the
contrast of vortex cores could serve as a sensitive thermometer
for a condensate in the regime for which the temperature is less
than the chemical potential and other methods of thermometry
become unreliable. In Sec.~V we discuss our preliminary efforts to
realize this vision. We are able to see an effect, but we have not
yet been able to extend this measurement technique below the usual
limits.
\par
The rest of Sec.~I discusses experimental issues common to all the
results of this paper

\subsection{A. Experiment}%\label{Exp}

Our experiment begins with a magnetically trapped cloud containing
greater than $10^{7}$ $^{87}$Rb atoms in the
$|F=1,m_{F}=-1\rangle$ hyperfine ground state, cooled to a
temperature roughly three times the critical temperature ($T_{c}$)
for Bose condensation. Using a TOP trap we confine these atoms in
an axially symmetric, oblate and harmonic potential with trapping
frequencies ($\{\omega_{\rho},\omega_{z}\}=2\pi\{7,13\}$Hz) with
axis of symmetry along the vertical ('z') axis. Rotation is
generated in the thermal cloud by resonantly coupling to a
scissors mode of the cloud~\cite{SandroSc,FootSc}. To do this we
gradually apply an elliptical deformation to the magnetic trapping
potential by distorting the amplitude of the rotating TOP field in
time. The resulting distorted trap has roughly similar average
radial-trapping frequency but an ellipticity~\cite{EllipEqn} in
the horizontal plane of 25\%. The uncondensed cloud is held in
this trap for 5~seconds while any excitations die out. At this
point the angle of the major axis of the elliptically distorted
trapping field is jumped quickly by 45~degrees in the horizontal
plane to generate the initial conditions of the scissors mode.
From here the cloud is allowed to evolve for 155~ms, or roughly
one quarter period of the resulting scissors mode, at which point
we transfer the cloud to a radially symmetric trap. Essentially we
have caught the scissors oscillation between turning points where
all the initial linear velocity has turned into rotational
velocity. Using this method we can generate a cloud rotating at
roughly half the radial trap frequency, with minimal heating. By
lowering the amplitude of the trap distortion or the angle by
which the trap is jumped, we can easily generate more slowly
rotating clouds as well.
\par
At this point we begin a second phase of rf evaporation, but this
time we evaporate in one dimension along the axis of rotation
\cite{Paul}. The motivation for this seemingly inefficient
technique is that the 1D evaporation allows us to remove energy
from the z axis of the condensate without preferentially removing
high angular momentum atoms. Lowering the energy per particle
without lowering the angular momentum per particle accelerates the
cloud rotation rate $\Omega$. To perform this 1D evaporation, we
adiabatically ramp to a prolate geometry
($\{\omega_{\rho},\omega_{z}\}=2\pi\{8.3,5.3\}$Hz) where the rest
of the experiment is carried out.
\par
Reaching significant rotation rates by the end of evaporation
requires that the lifetime of the thermal cloud's angular momentum
be comparable to the evaporation time. The nearly one-dimensional
nature of the evaporation together with the low average trap
frequencies makes cooling to BEC in the prolate trap very slow (2
minutes). We obtain angular momentum lifetimes this long by
shimming the TOP trap's rotating bias field to cancel the few
percent azimuthal trap asymmetry that exists despite careful
construction. With this technique we suppress the azimuthal trap
ellipticity to less than one part in a thousand.
\par
After the evaporation we have a condensate with as many as 4.5
million atoms and rotation rates from
$\Omega=(0-0.975)\omega_{\rho}$, with no observable thermal cloud.
Rotation can be accurately determined by comparing the condensate
aspect ratio $\lambda$ (defined as the axial Thomas-Fermi radius
over the radial Thomas-Fermi radius $R_{z}/R_{\rho}$) to the trap
aspect ratio $\lambda_{0}\equiv(\omega_{\rho}/\omega_{z})$, and
using the now standard relation
\begin{equation}\label{rotation}
    \Omega/\omega_{\rho}=\sqrt{(1-(\lambda/\lambda_{0})^{2})}~.
\end{equation}

\subsection{B. Expansion}%\label{expansion}

The clouds described here typically contain between 1-200
vortices, each of which is too small to be observed in-trap but
can be seen after expansion of the cloud.  Our expansion
technique, while not completely unique, is unusual enough to
warrant a description. For the experiments presented here and
elsewhere, we need a large radial expansion to make sure that the
vortex cores are large compared to our imaging resolution.
Additionally we need to suppress the axial expansion in order to
preserve certain length scales in the condensate as discussed
below. Clearly, given our low and nearly isotropic trap
frequencies, the usual expansion technique of shutting off the
magnetic field and dropping the cloud would not meet such
requirements. The solution, which has been demonstrated by other
groups \cite{Heather}, is to perform an anti-trapped expansion.
Rather than simply shutting off the trapping potential, we invert
the trap in the radial direction so that the cloud is actively
pulled apart. Simultaneously, the magnetic field gradient in the
vertical direction is used to support against gravity. Because it
has generated some interest we describe this technique in the
following paragraph in what otherwise might be excessive detail.
\par
The expansion is achieved in several steps that take place in
rapid succession.  First we employ a microwave adiabatic rapid
passage technique (ARP) to transfer the atoms from the weak-field
seeking $|F=1,m_{F}=-1\rangle$ state to the strong-field seeking
$|F=2,m_{F}=-1\rangle$ state.  The microwave field employed is
powerful enough to perform the transfer in $10 \mu$s but, as will
be discussed shortly, we often take as long as $300 \mu$s for this
transfer. After transfer to the anti-trapped state, the cloud
still sits in its original position below the quadrupole zero,
which means that both gravity and the magnetic field are acting to
pull it downward.  To counter this force a downward uniform
vertical magnetic field is added to pull the quadrupole zero below
the condensate so that the magnetic field gradient again cancels
gravity. The field is applied within $10 \mu s$, fast compared to
relevant time scales. In this manner the cloud is again supported
against gravity.  To reduce curvature in the z direction, the TOP
trap's rotating bias field is turned off leaving only the linear
magnetic gradient of the quadrupole field. This gradient is tuned
slightly to cancel gravity.
\par
Using this technique, we are able to radially expand the cloud by
more than a factor of 10 while, at the same time, seeing less than
a factor of two axial expansion.  Unfortunately even this much
axial expansion is unacceptable in some cases.  In the limit of
adiabatic expansion, this factor of two decrease in condensate
density would lead to an additional $\sqrt{2}$ increase in healing
length during expansion. Thus, features that scale with healing
length, such as vortex core radius in the slow rotation limit,
would become distorted. The effect of axial expansion on vortex
size was first noted by Dalfovo and Modugno \cite{Modugno}.
\par
To suppress the axial expansion, we give the condensate an initial
inward or compressional impulse along the axial direction. This is
done by slowing down the rate at which we transfer the atoms into
the anti-trapped state. The configuration of the ARP is such that
it transfers atoms at the top of the cloud first and moves down
through the cloud at a linear rate.  These upper atoms are then
pulled downward with a force of 2g (gravity plus magnetic
potential), thus giving them an initial inward impulse. Finally,
the ARP sweep passes resonantly through the lowest atoms in the
cloud: they, too, feel a downward acceleration but the axial
magnetic field gradient is reversed before they can accumulate
much downward velocity.  On average the cloud experiences a
downward impulse, but also an axial inward impulse. Normally the
ARP happens much too fast for the effect to be observable but when
the transfer time is slowed to $200-300 \mu$s the effect is enough
to cause the cloud to compress axially by 10-40\% for the first
quarter of the radial expansion duration.  The cloud then expands
back to its original axial size by the end of the radial
expansion.
\par
\begin{figure}
\begin{center}

\psfig{figure=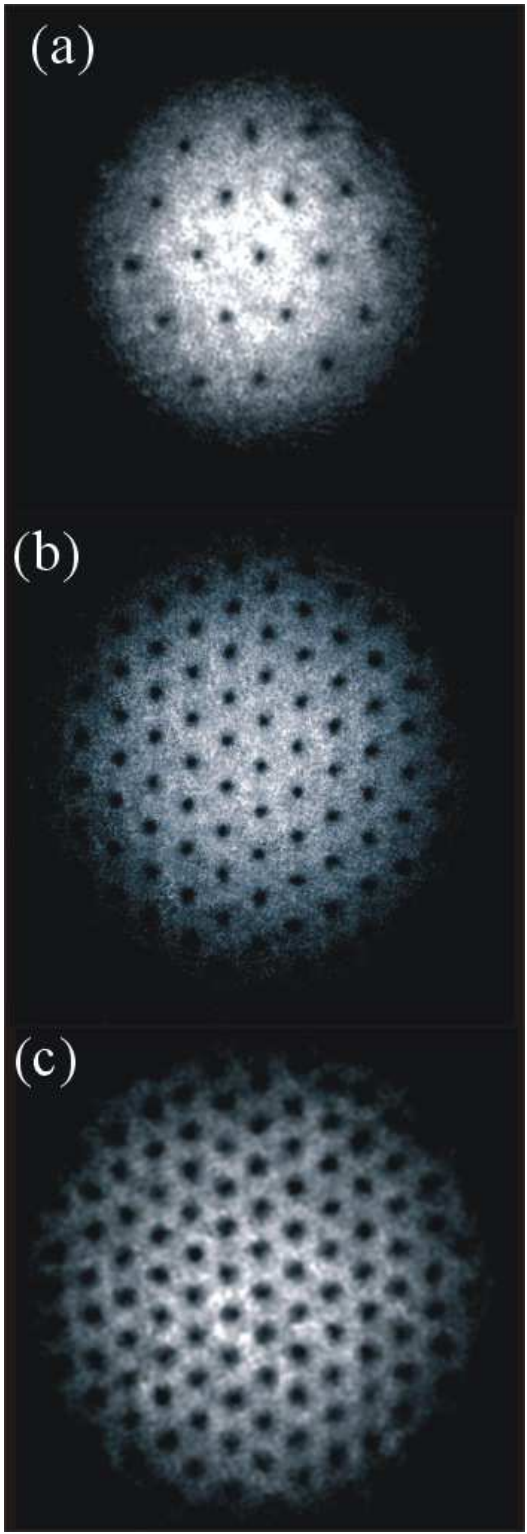,width=.8 \linewidth,clip=} % 0.5
\end{center}
\caption {Examples of the condensates used in the experiment
viewed after expansion. Image (a) is a slowly rotating condensate.
Images (b) and (c) are of rapidly-rotating condensates with
similar in-trap conditions. They differ only in that (c) was
allowed to expand axially during the anti-trapped expansion.  The
effect on the vortex core size is visible by eye. Images (a) and
(b) are of the regularity required for the nearest-neighbor
lattice spacing measurements. } \label{Vimages}
\end{figure}

Despite our best efforts to null out axial expansion, we observe
that the cloud experiences somewhere between 20\% axial
compression to 20\% axial expansion at the time of the image,
which should be, at most, a 10\% systematic error on measured
vortex core radius. The overall effect of axial expansion can be
seen in Fig.~\ref{Vimages}, where image (b) and (c) are similar
condensates and differ primarily in that (c) has undergone a
factor of 3 in axial expansion while in (b) axial expansion has
been suppressed. The effect on the vortex core size is clearly
visible.
\par

Because almost all the data presented in this paper is extracted
from images acquired after the condensate expands, it is worth
discussing the effect of radial expansion on the density structure
in the cloud. In the Thomas-Fermi limit, it is easy to show that
the anti-trapped expansion in a parabolic trap, combined with the
mean-field and centrifugally driven expansion of the rotating
cloud, leads to a simple scaling of the linear size \cite{Castin}
of the smoothed, inverted-parabolic density envelope. As
$R_{\rho}$ increases, what happens to the vortex-core size?  There
are two limits that are easy to understand.  In a purely 2D
expansion (in which the axial size remains constant), the density
at any spot in the condensate comoving with the expansion goes as
$1/R_{\rho}^{2}$, and the local healing length $\xi$ then
increases over time linearly with the increase in $R_{\rho}$. In
equilibrium, the vortex core size scales linearly with $\xi$. The
time scale for the vortex core size to adjust is given by
$\hbar/\mu$ where $\mu$ is the chemical potential. In the limit
(which holds early in the expansion process) where the fractional
change in $R_{\rho}$ is small in a time $\hbar/\mu$, the vortex
core can adiabatically adjust to the increase in $\xi$, and the
ratio of the core-size to $R_{\rho}$ should remain fixed as the
cloud expands.
\par
In the opposite limit, which applies when $R_{\rho}$ expands very
rapidly compared to $\hbar/\mu$, the inverted parabolic potential
dominates the dynamics, and every point in the cloud expands
radially outward at a rate proportional to its distance from the
cloud center. In this limit, the ``fabric of the universe" is
simply stretched outward, and all density features, including
vortex core size expand at the same fractional rate. Again, the
ratio of core-size to $R_{\rho}$ should remain fixed.
\par
So in the two extreme limits, the ration of vortex core size to
$R_{\rho}$ (and other density features, such as nearest-neighbor
vortex separation) remains fixed. It is reasonable then to assume
that this behavior will be true in general in the intermediate
regime between the two expansion rates. Extensive numerical
simulations were performed to validate this assumption.
\par
On a separate note, the vortex lattice spacing, unlike vortex core
size, is largely unaffected by the presence of axial expansion.
However, we see some indication that deep in the lowest Landau
level large axial expansions can affect vortex spacing as well as
size.
\par
Once expanded, the cloud is imaged along the vertical direction,
and data is extracted by fitting the integrated (along the line of
sight) condensate density profile with a Thomas-Fermi
distribution. We then subtract this fit from the image and easily
fit the remaining vortex-core residuals with individual 2D
Gaussian profiles to determine the core centers and radii. For the
purpose of this paper the vortex radius $r_{v}$ is defined to be
the RMS radius of the 2D Gaussian that we fit to the core. For
clarification, the RMS radius of a 2D Gaussian would be 0.60 times
its FWHM. Condensate and vortex fits can be performed iteratively
to reduce error.
\par
Before each expanded image we also take a horizontal,
nondestructive, in-trap image of the cloud immediately before
expansion.  From this image, rotation rate and atom number are
determined. Length scales in the expanded cloud can be scaled back
to in-trap values by dividing by the radial expansion factor,
defined as $R_{\rho}(expanded)/R_{\rho}(in-trap)$.

\subsection{C. Numerical studies}%\label{numerics}

The numerical studies discussed in this paper are done by setting
up an initial in-trap condensate wave function with a given N and
$\Omega$ on a 2048x2048 lattice and then relaxing this wave
function by propagating the Gross-Pitaevskii equation in imaginary
time.  All work shown in Sec.~III~\small{B}\normalsize~and
III~\small{C}\normalsize~is done in 2D. Additionally a radially
symmetric 3D simulation can be performed for a single vortex, as
is done in Sec. III~\small{A}\normalsize. Once the final wave
function is found, we convert to an atom density profile which can
be fit and analyzed in the same manner as the experimental data.

\section{II. The Lattice Constant}%\label{SecLatt}

At first sight, vortex lattices, such as the one seen in
Fig.~\ref{Vimages}(b), appear perfectly regular. However as noted
in the introduction, Sheehy and Radzihovsky~\cite{Dan} predict
that there should exist a small correction to the vortex density
in the condensate due to the condensate, density inhomogeneity.
One result from Ref. \cite{Dan} is that the areal density of
vortices is
\begin{equation}\label{vortexdensity}
n_{v}(\rho) = \frac{\Omega m}{\pi\hbar}-\frac{1}{2\pi
R_{\rho}^{2}(1-(\rho/R_{\rho})^2)^2}ln[\hbar/(2.718\,m
\Omega\xi^{2})]~,
\end{equation}
where m is the mass of rubidium and $\xi$ is the healing length
(calculated from the measured density). This equation can
conveniently be thought of as the rigid body rotation (first term)
plus the density inhomogeneity correction that reduces vortex
density. We compare to experimental measurements by converting
vortex density to a nearest-neighbor lattice spacing, conveniently
expressed in units of condensate radius
\begin{equation}\label{latticeconst}
    b(\rho) = \sqrt{2/(3^{1/2} n_{v}(\rho))}\frac{1}{R_{\rho}}~.
\end{equation}
\par
To study this lattice inhomogeneity effect experimentally, we
generate condensates with rotation rates of
$\Omega/\omega_{\rho}=0.5 -0.9$. To extract the vortex separation,
we expand the cloud by a factor of 10 in the radial direction
using the anti-trapped expansion technique. The condensate and
vortices are fit as described in Sec.~I~\small{B}\normalsize.~ The
nearest-neighbor separation for a given vortex is measured by
averaging the distance from the vortex center to the centers of
the six nearest vortices.  Because of low signal, vortices further
than 0.9 $R_{\rho}$ from the condensate center are disregarded.
Any remaining vortex with fewer than six nearest neighbors (i.e.,
a vortex in the outer ring) is used as a neighbor to other
vortices but is not itself included in the final data. Obviously
using the six nearest neighbors assumes a triangular lattice
structure, so before fitting, each image is checked for defects in
the lattice. Any image exhibiting broken lattice planes is not
considered. Once the nearest-neighbor separation is measured, it
is normalized by the expanded condensate radius to compare to Eq.
\ref{latticeconst}. For this comparison, $R_{\rho}$, $\Omega$, and
$\xi$ are measured or calculated from an in-trap image. To improve
the theory fit, we allowed $R_{\rho}$ to be an adjustable theory
parameter, but, in each case, the theory fit value for $R_{\rho}$
was within 5\% of the experimentally determined value of
$R_{\rho}$.  This $<5\%$ difference is within the calibration
uncertainty for such a measurement. Noise is suppressed by taking
an azimuthal average of the lattice spacing data.  Due to the
discrete nature of the vortices this is equivalent to binning the
lattice-spacing data by radial displacement of the vortex from the
condensate center.

\begin{figure}
\begin{center}
\psfig{figure=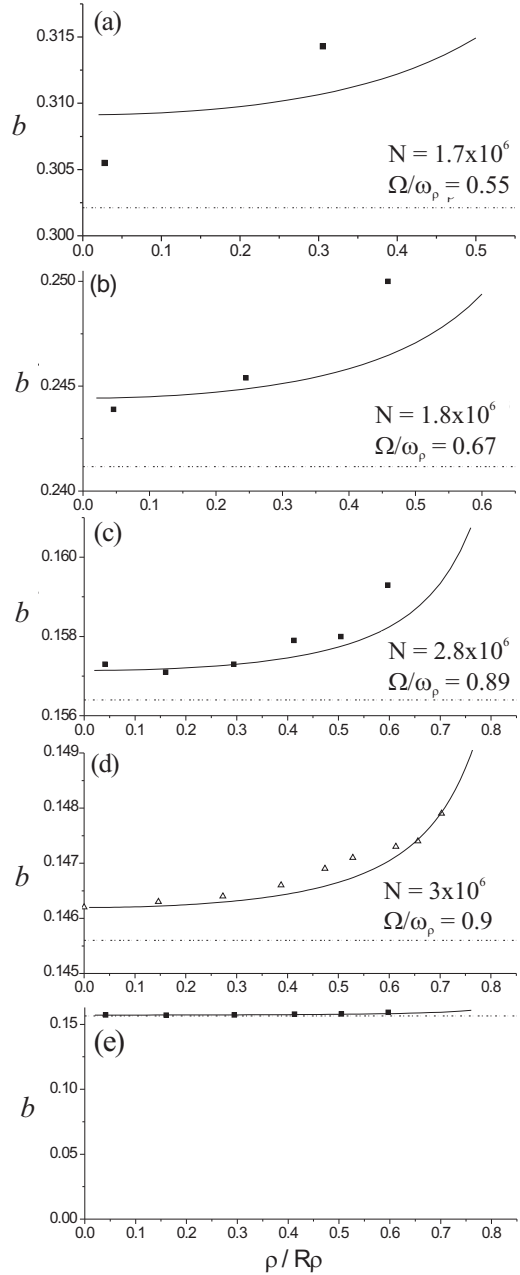,width=.9\linewidth,clip=} % 0.5
\end{center}
\caption {Measured and binned lattice spacings as a function of
radial position $\rho$.  The solid curve is the theory result
(Eq.~(\ref{latticeconst})) of Sheehy and Radzihovsky \cite{Dan}.
The rigid-body--rotation rate lattice spacing is also plotted for
comparison (dashed line). Plots (a-c) are experimental data with
increasing rotation. Plot (a) and (c) are data taken from the
condensate in Figs.~\ref{Vimages}(a) and (b), respectively. Plot
(d) is the same effect observed in the numerical data. One can see
that theory and experiment show a similar dependence on radial
position and that the fractional amplitude of the density
inhomogeneity effect is suppressed at higher rotation.  Plot (e)
is the data in (c) plotted without suppressing the zero. The
vortex lattice spacing changes less than 2\% over a region in
which the atom density varies by 35\%.} \label{SheehyPlot}
\end{figure}

\par
Figure~\ref{SheehyPlot} shows a comparison to theory for three
physical condensates and one numerically generated condensate
density profile. Figure~\ref{SheehyPlot}(a) is data taken from the
condensate in Fig.~\ref{Vimages}(a).  The two points shown
correspond to the measured vortex density for the center vortex
(first point) and the average vortex density for the first ring of
vortices. Also plotted is Eq.~\ref{latticeconst} (solid line) and
the expected nearest-neighbor lattice spacing for rigid body
rotation (dashed line). The imperfect fit may be partly due to the
discrete nature of the data, vis-a-vis a continuum
theory~\cite{Dan}. Plots (b) and (c) are condensates with
increasing rotation rates where (c) is taken from
image~\ref{Vimages}(b). Plot (d) is a comparison with numerical
data prepared with parameters similar to the experimental
situation in (c). Figure~\ref{SheehyPlot}(e) is the same data in
Fig.~\ref{SheehyPlot}(c) but plotted without the suppressed zero
to emphasize the smallness of the position dependant effect. The
areal density of vortices is constant to 2\% over a region that
experiences an atom-density variation of 35\%.

\section{III. Vortex Cores}%\label{SecCoreSize}
\subsection{A. Core size}%\label{coresize}

The other defining length scale of the vortex lattice is the core
radius.  Here we study the core radius in the Thomas-Fermi regime
(as opposed to the lowest Landau level regime, described later)
where it should scale with the healing length. A theoretical value
for the vortex core radius was generated by performing a numerical
simulation for a 3D BEC containing an isolated vortex and
comparing the fitted radius of this vortex to the corresponding
healing length. Fitting the simulation in the same manner that we
later treat the experimental data (described in
Sec.~I~\small{B}\normalsize) we obtain an expression for the core
radius of
\begin{equation}\label{rvEqn}
    r_{v}=1.94\times\xi~,
\end{equation}
with healing length $\xi$=$(8\pi n a_{sc})^{-1/2}$, where $a_{sc}$
is the scattering length and $n$ is the density-weighted atom
density. For the data presented, $n$ is determined from the
in-trap image before expansion.
\par
Core size measurements and fractional core area (discussed in the
next subsection) measurements require considerable attention to
detail. In pursuing these measurements, we find that nearly
everything~--- from focusing issues, to lensing due to off
resonant imaging light, to even imperfect atom transfer into the
anti-trapped state before expansion~--- can lead to an
overestimation of the vortex core size. By far the biggest
potential systematic error in our system is axial expansion,
which, as noted in Sec.~I~\small{B}\normalsize, requires careful
attention.

\begin{figure}
\begin{center}
\psfig{figure=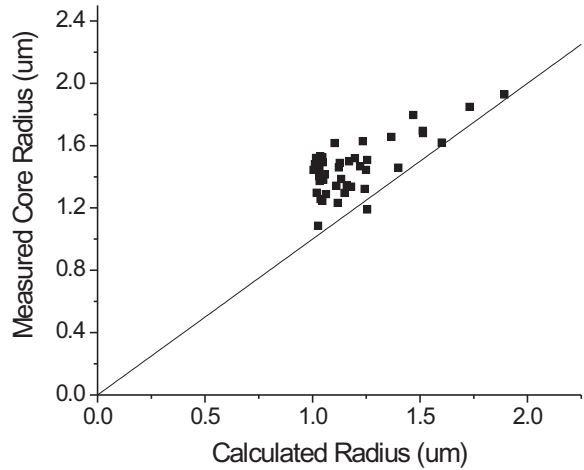,width=1\linewidth,clip=}
\end{center}
\caption {Comparison of measured core radii with the Thomas-Fermi
prediction (Eq.~(\ref{rvEqn})) represented by the solid line.
Black squares are the core size in expansion scaled by the radial
expansion of the condensate so that they correspond to the in-trap
values.  Data shows reasonable agreement with theory.  The fact
that the measured core size is consistently larger is likely due
to the fact that nearly all our imaging systematics lead to an
overestimation of the vortex core size.} \label{coresize}
\end{figure}

\par
A range of core sizes is achieved by varying the initial number of
atoms loaded into the magnetic trap prior to evaporation. To avoid
the core size saturation effect, due to high condensate
rotation~\cite{BaymLLL}, we consider only clouds with
$\Gamma_{LLL}>10$, where $\Gamma_{LLL}\equiv\mu/(2 \hbar\Omega)$
is the LLL parameter and $\mu$ is the chemical potential. This
ratio of chemical potential to rotation approaches unity as we
enter the LLL regime, while at values of 10 or greater we should
be firmly in the Thomas-Fermi regime. In practice this requires
only that we keep the condensate rotation rate low. Core size is
measured by fitting the expanded image with a Thomas-Fermi profile
and each core with a 2D Gaussian. For Fig.~\ref{coresize} the
measured core radius in expansion is scaled back to the
corresponding in-trap value using the radial expansion factor
discussed in Sec.~I~\small{B}\normalsize. To reduce scatter we
consider only vortices located less than half a condensate radius
out from the center. Additionally we find that some vortices
appear to have some excitation or bending which leads to a poor
fit. To filter these out we consider only vortices that have a
contrast greater than 0.6. Here contrast is defined, with respect
to the integrated (along the line of sight) condensate profile, as
the peak of the ``missing" column density at the vortex position
divided by the smoothed Thomas-Fermi profile at the same position.
On average, about 30\% of visible vortices fulfill all the
criteria for being used in the core size measurement, although
this number can vary wildly (10\%-100\%) depending on vortex
number and position.

\par
From Fig.~\ref{coresize} we can see that the data and the
Thomas-Fermi theory agree reasonably well.  The data do seem to be
slightly above the theory value, but we are hesitant to make too
much of this because, as noted before, there are many systematic
errors that tend to bias the data toward larger core size.
Measurement is easier and the agreement better, on the low
density, large-core side of the graph.
\par
At an early stage in this work, we speculated that the mean-field
Gross-Pitaevskii equation might not give a good quantitative
description of vortex core size because the core size is
particularly sensitive to the healing length $\xi$. At our highest
densities, while the gas is nominally dilute ($na_{sc}^{3} <
10^{-5}$, where $a_{sc}$ is the interatomic scattering length),
the mean interatomic distance $n^{-1/3}$ is only a factor of $1.5$
less than $\xi$. Our data, however, are ambiguous with respect to
the accuracy of the Gross-Pitaevskii equation for predicting
vortex core sizes at our highest densities. The roughly 25\%
discrepancy between our measurements and the mean-field prediction
shown at the smaller radii in Fig.~\ref{coresize} is comparable to
possible systematic errors in our measurements of the smaller
cores that exist at higher densities. In retrospect, our
experimental design is such that we may be unable to see a
mean-field failure even if one were to exist. During the radial
expansion, the density drops. Thus the accuracy of the mean-field
approximation is likely to improve significantly during the
expansion. Our anti-trapped expansion, while more rapid than a
conventional ballistic expansion, is still slow compared to the
rate at which a vortex can adiabatically relax its
radius~\cite{BaymLLL} (approximately $\mu/\hbar$). Any
non-mean-field corrections to the vortex core size will likely
relax away before the cores have expanded to be large enough for
us to reliably image them~\cite{2d}.

\subsection{B. Fractional core area}%\label{FArea}

As noted in the introduction, the fractional condensate area
occupied by the vortex cores is also a quantity that has been of
much theoretical interest~\cite{BaymLLL,FederNv,FetterLLL} and has
been experimentally studied previously \cite{JilaLLL}. It is
argued by Baym and Pethick~\cite{BaymLLL} that the fractional core
area reaches a limiting value as one enters the LLL regime. The
corollary of this argument is that fractional core area is a
reasonable way to monitor the transition to the lowest Landau
level regime. We examine this transition with numerical work,
which we can to push further into this regime than we can achieve
experimentally. Additionally we examine some of the systematic
errors that can affect the experimental data. To this end
numerical calculations were performed as previously described, for
$3\times10^{6}$,$5\times10^{5}$,and $1\times10^{5}$ atoms, and for
rotations ranging from $\Omega/\omega_{\rho}=$0.15 to 0.998. For
the experimental data, actual condensates were generated over a
similar range with $\Omega/\omega_{\rho}=$0.15 to 0.98 and
N=$4-50\times10^{5}$. The numerical data as well as the
experimental data are fit in the same manner as described in
Sec.~I~\small{B}\normalsize.
\par
\begin{figure}
\begin{center}
\psfig{figure=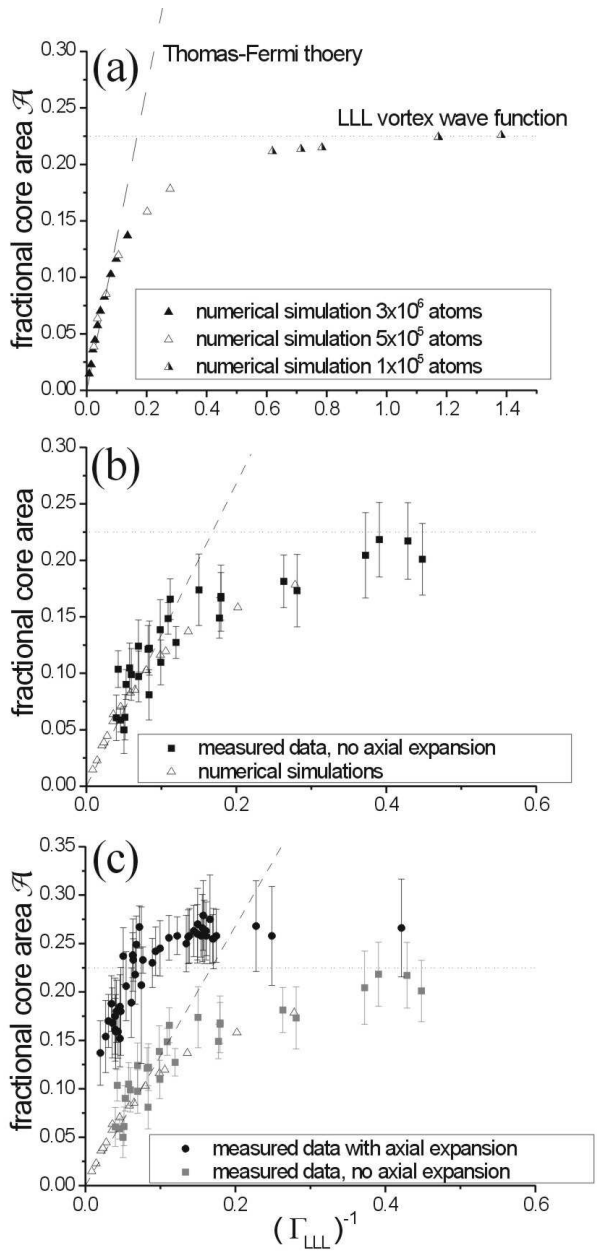,width=1\linewidth,clip=} % 0.5
\end{center}
\caption {Fractional condensate area occupied by vortex cores
($\mathcal{A}$) as a function of
$\Gamma_{LLL}^{-1}=2\hbar\Omega/\mu$, the inverse lowest Landau
level parameter.  Plot (a) shows a smooth transition in the
numerical data from the Thomas-Fermi limit where $\mathcal{A}$ is
linear in $\Gamma_{LLL}^{-1}$ to the LLL limit where $\mathcal{A}$
saturates.  Here the Thomas-Fermi theory is represented by the
dashed line and the LLL limit by the dotted line. Plot (b) is a
comparison of the numerical data to experimental data. Plot (c)
demonstrates the effect on the experimental measurement of
allowing the condensate to expand axially during the expansion
process.} \label{FracArea}
\end{figure}

We define the fractional area $\mathcal{A}$ occupied by the
vortices to be $\mathcal{A}=n_{v}\pi$r$_{v}^{2}$, where $n_{v}$ is
the areal density of vortices and r$_{v}^{2}$ is defined in
Eq.~\ref{rvEqn}. Ignoring density inhomogeneity effects, in the
limit of many vortices, the expected vortex density $n_{v}$ is $
m\Omega/(\pi\hbar)$. The resulting prediction for $\mathcal{A}$
can be expressed as $\mathcal{A}=1.34\times\Gamma_{LLL}^{-1}$.
This value exceeds unity for $\Gamma_{LLL}<1.34$, which has led to
the prediction that vortices should merge as the condensate enters
the LLL regime. An alternate treatment from Baym and Pethick
\cite{BaymLLL} predicts that $\mathcal{A}$ saturates at 0.225 as
the vortices go from a Thomas-Fermi profile to the profile of a
LLL wave function. Our numerical data for $\mathcal{A}$, together
with experimental points, are plotted in Fig.~\ref{FracArea}
(a,b). For $\Gamma_{LLL}^{-1}<0.1$, the data agree reasonably well
with the Thomas-Fermi result. For larger $\Gamma_{LLL}^{-1}$, the
data clearly show a smooth transition to the LLL regime, and a
saturation of $\mathcal{A}$ at the LLL limit.
\par

The experimental data in Fig.~\ref{FracArea}(b) tend to lie above
the numerical data.  This is likely due to our tendency to
overestimate the core size, as discussed previously in
Sec.~III~\small{A}\normalsize. Figure~\ref{FracArea}(c)
demonstrates the dangers of axial expansion in this measurement.
For the data presented the condensate undergoes a factor of 2-3 in
axial expansion, and we see a corresponding increase in
$\mathcal{A}$. This clearly illustrates the importance in
suppressing axial expansion for these measurements. It is
interesting to note that with our rapid axial expansion the
fractional core area can overshoot the LLL limit, which in
principle should still be valid in the limit of adiabatic
expansion.

\subsection{C. Core density profile }%\label{coreprofile}

\begin{figure}
\begin{center}
\psfig{figure=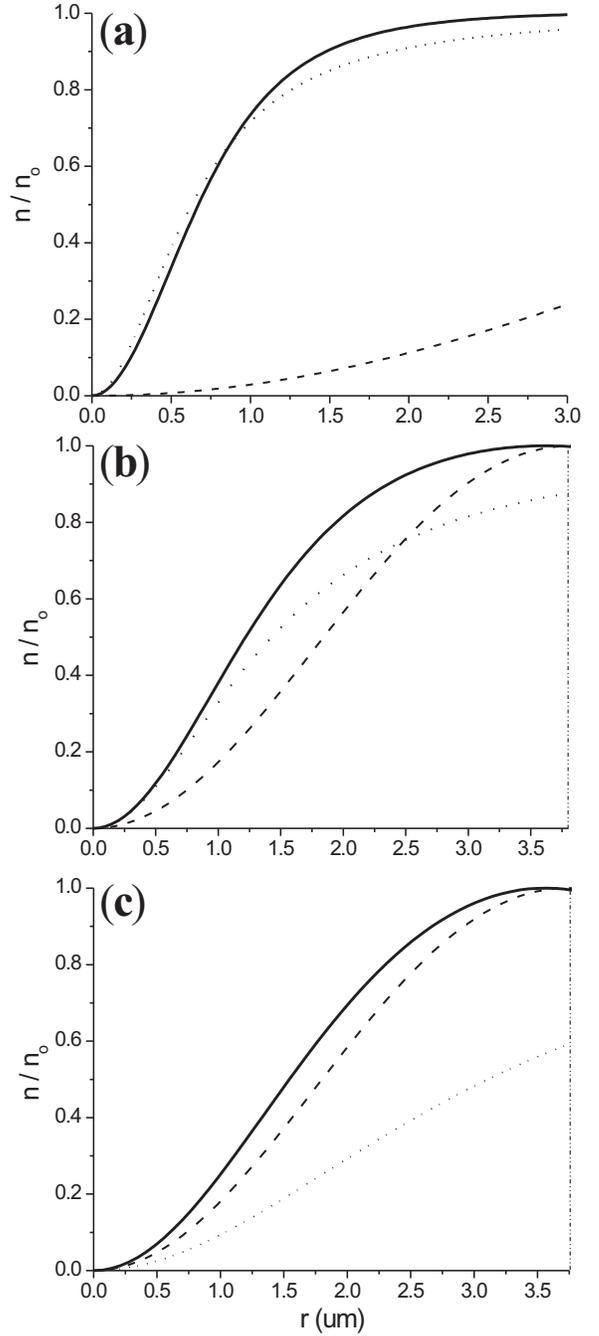,width=1\linewidth,clip=} % 0.5
\end{center}
\caption {Numerically generated vortex core density profiles
approaching the lowest Landau level regime.  Solid lines represent
the numerical result for (a) $\Gamma_{LLL}=117, (b)
\Gamma_{LLL}=3.6, (c) \Gamma_{LLL}=0.72$.  The dashed line is the
expected profile for a LLL wave function \cite{BaymLLL} given the
condensate rotation.  The dotted line is the expected vortex form
in the Thomas-Fermi limit\cite{FetterCores} given the condensate
density. The vertical lines in figure (b) and (c) designate the
edge of the Wigner-Seitz unit cell. As $\Gamma_{LLL}$ decreases,
one can see a clear transition from the interaction-dominated
Thomas-Fermi regime to a LLL function where kinetic energy
concerns and the vortex core spacing dictate the shape and size of
the vortex.} \label{coreshape}
\end{figure}

We can also observe the transition to the LLL regime in the
numerical data by examining the shape of the condensate vortex
cores. In the Thomas-Fermi regime, the vortex-core density profile
is well described by the form $n(r)=
(r/\sqrt{2\xi^{2}+r^{2}})^{2}$\cite{FetterCores} where $r$ is
measured from the vortex center. Alternatively in the LLL regime,
the core is no longer dictated by the interactions but rather by
kinetic energy considerations.  In this case, within the
Wigner-Seitz unit cell, the vortex is thought \cite{BaymLLL} to
have a simple oscillator p-state structure $n(r)= ((C r/b)\cdot
exp[-r^{2}/2 l^{2}])^{2}$ for $0\leq r\leq l$, where C is a
normalization constant, and $l$ is the radius of the Wigner-Seitz
unit cell and is related to the nearest-neighbor lattice spacing
$b$ by $l=(\sqrt{3}/2\pi)^{1/2}b$. Figure~\ref{coreshape} is a
comparison of the central vortex, in three numerically generated
condensates, to both Thomas-Fermi and LLL predicted core shapes.
The simulation for Fig.~\ref{coreshape}(a) was performed for
$3\times 10^{6}$ atoms and $\Omega/\omega_{\rho}=0.15$ and is well
inside the Thomas-Fermi regime ($\Gamma_{LLL}=117$). Here the
density profile of the numerical data (solid line) seems to fit
quite well to the Thomas-Fermi vortex form (dotted line), but the
LLL form is a poor description of the vortex core (dashed line).
The simulation for Fig.~\ref{coreshape}(b) was performed for
$5\times 10^{5}$ atoms, $\Omega/\omega_{\rho}=0.95$ and
$\Gamma_{LLL}=3.6$. One can see from Fig.~\ref{FracArea} that this
is in the transition region. Not surprisingly both vortex forms
fit about equally well. In Figs.~\ref{coreshape}(b) and
Fig.~\ref{coreshape}(c), the vertical line represents the edge of
the Wigner-Seitz unit cell at $r=l$. The simulation for
Fig.~\ref{coreshape}(c) was performed for $1\times 10^{5}$ atoms,
$\Omega/\omega_{\rho}=0.998$ and $\Gamma_{LLL}=0.72$. One can see
that LLL is a much better description of the vortex.
\par

On a separate but interesting note, even this far into the LLL
regime, our numerical solution of the GP equation shows that the
radial profile of the overall smoothed condensate fits much better
to a parabola than to the Gaussian that was originally predicted
\cite{HoLLL}. The reason the Gaussian-density-profile prediction
fails to pan out can be extrapolated from data presented earlier
in this paper. The density-profile prediction for the radial
profile in the LLL arose from an elegant argument that was based
on an assumption that the vortex nodes were on a perfect
triangular lattice.  As was originally pointed out to us by A. H.
MacDonald \cite{allan} and has been the subject of two recent
theoretical works \cite{BaymLatt,ReadLatt}, a slight radially
dependant perturbation in the areal density of vortices is enough
to convert a Gaussian density distribution into an inverted
parabola. The analytic description of this perturbation in
\cite{BaymLatt} (calculated in the LLL) bears a striking
resemblance to the one measured in Sec.~II in the Thomas-Fermi
regime and also to the analytic form \cite{Dan} calculated in the
Thomas-Fermi regime.  The surprising result of this perturbation
in the areal density of vortices is that one of the most striking
features of the Thomas-Fermi regime, the parabolic Thomas-Fermi
density profile, still exists in the LLL regime where the
condensate kinetic energy is clearly non-negligible compared to
interaction energy.

\section{IV. Rotational suppression of the quantum degeneracy
temperature}%\label{SecDegeneracy}

So far in this article, condensate rotation has been considered in
terms of its discrete vortex elements and their dynamics. A more
global effect of condensate rotation is to suppress the quantum
degeneracy temperature, $T_{c}$, for a fixed number of trapped
particles, $N$ \cite{SandroTc}. While superficially similar to
magnetic-field suppression of $T_{c}$ in superconductors, this
effect is distinct in that it is not a many-body effect but rather
a suppression of atom density due to a rotationally weakened
trapping potential.
\par

\begin{figure}
\begin{center}
\psfig{figure=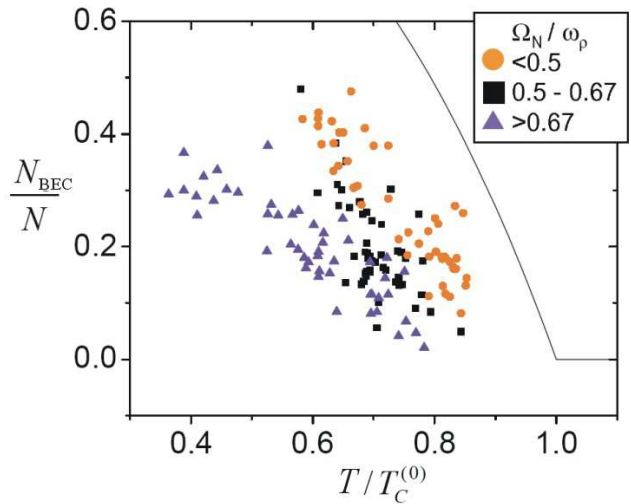,width=1\linewidth,clip=}
\end{center}
\caption {(Color online) Condensate fraction versus temperature
for various sample rotation rates.  Condensate number, $N_{BEC}$,
and total number, $N$, are obtained from fits to cloud images.
The temperature $T$ is extracted from the vertical width of the
thermal cloud while the centrifugal distortion of the thermal
cloud's aspect ratio yields its rotation $\Omega_{N}$. The
temperature is scaled by the static critical temperature
$T_{c}^{(0)}$ for an ideal gas. The data has been grouped
according to three different ranges of $\Omega_{N}$. The solid
line is the theoretical dependance expected for a static ideal
gas.} \label{TcPaul}
\end{figure}

\begin{figure}
\begin{center}
\psfig{figure=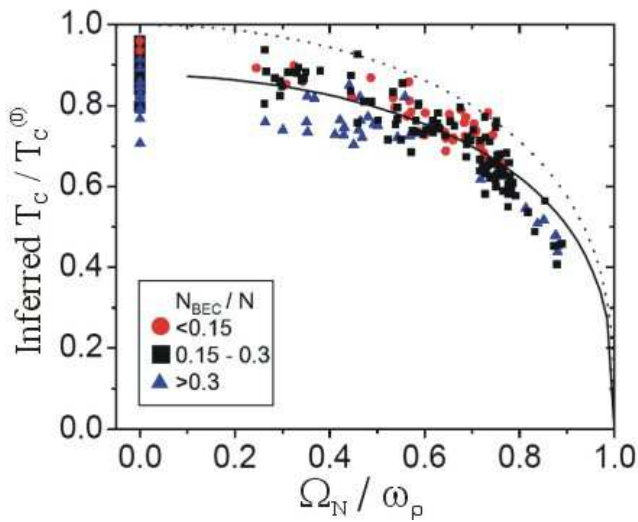,width=1\linewidth,clip=}
\end{center}
\caption {(Color online) Inferred critical temperature $T_{c}$,
scaled by the non-rotating expectation $T_{c}^{(0)}$ for an ideal
gas, as a function of thermal gas rotation $\Omega_{N}$. For each
data point, $T_{c}$ is inferred from the measured condensate
fraction and temperature of a sample. The total number of atoms is
used to obtain $T_{c}^{(0)}$. The data has been grouped according
to three different condensate fractions as shown. A set of static
points, where the thermal cloud is not stirred before evaporation,
are deliberately plotted at zero rotation. Otherwise, the rotation
rate $\Omega_{N}$ is obtained from the thermal cloud's aspect
ratio. The rotating data has been cropped at a minimum threshold
$0.2 < \Omega_{N}/\omega_{\rho}$ to avoid imaginary rotation
values arising from noise in near-static aspect ratios. The dotted
line is the theoretical expectation according to Eq.~(\ref{TcEqn})
with $T_{c}^{(0)}$ as for an ideal gas. The solid line is a fit of
the data to Eq.~(\ref{TcEqn}) with an arbitrary overall scaling of
the vertical axis. The fit result is equivalent to assuming an
effective $T_{c}^{(0)}$ that is $87\pm1\%$ of the ideal gas value.
This 13\% shift is consistent with what one would expect from
atom-atom interactions \cite{AspectTc}.} \label{Tc2Paul}
\end{figure}

One can define a static critical temperature, $T_{c}^{(0)}$ that
applies in the case of no rotation. For a non-interacting gas in a
harmonic potential, $T_{c}^{(0)} = 0.94\hbar\omega_{ho}N^{1/3}$,
where $N$ is the number of atoms in the sample and $\omega_{ho} =
(\omega_{\rho}^{2} \omega_{z})^{1/3}$ is the average trap
frequency. Rotational suppression of $T_{c}$ may be accounted for
by the centrifugal weakening of the radial harmonic confinement,
$\tilde{\omega}_{\rho}=\omega_{\rho}\sqrt{1-\Omega^{2}/\omega_{\rho}^{2}}$,
which leads to a reduction in the thermal gas density $n_{th}$
relative to the non-rotating case. As a result, for fixed N, a
lower temperature must be reached before the phase space density
$(\sim n_{th}/T^{3/2})$ is sufficiently high to bring about the
BEC transition. The expression for $T_{c}$ as a function of the
rigid-body rotation rate $\Omega$ of the gas sample is
\begin{equation}\label{TcEqn}
    T_{c}=T_{c}^{(0)}(1-\Omega^{2}/\omega_{\rho}^{2})^{1/3}~.
\end{equation}
In Fig. \ref{TcPaul}, experimental data have been used to plot
condensate fraction versus temperature for three different ranges
of sample rotation rate. All quantities are obtained from fits to
the non-destructive, in-trap images of the trapped gas clouds. As
usual, the rotation rate has been assessed from the changing
aspect ratio of the thermal cloud according to Eq.~\ref{rotation}.
It is qualitatively clear from Fig. \ref{TcPaul} that a lower
temperature is required at higher rotation rates to reach a given
condensate fraction. For each data point, the temperature, $T$,
and condensate fraction, $N_{BEC}/N$ (where $N_{BEC}$ is the
number in the condensate) can be used together to infer a value
for the critical temperature of the sample using the formula
$T_{c}=[1-N_{BEC}/N]^{-1/3}T$. To remove shot-to-shot variation in
$T_{c}$ due to atom-number fluctuations, the value of the inferred
transition temperature can be scaled by the static value
$T_{c}^{(0)}$ calculated from the measured atom number. The scaled
value of $T_{c}$ is plotted against rotation rate in
Fig.~\ref{Tc2Paul} for three ranges of temperature: ``hot,"
``medium" and ``cold," corresponding to three different ranges of
condensate fraction as given by the legend of Fig.~\ref{Tc2Paul}.
\par
The centrifugal suppression of $T_{c}$ is perhaps less interesting
than suppression due to many-body interactions. Nevertheless, from
a technical stand point, centrifugal effects are an important
consideration in the evaporative cooling and spin-up process.

\section{V. Core Contrast and Condensate Temperature}%\label{SecContrast}

Since the very first observations of dilute-gas BEC, the
temperature of the sample has been determined by imaging the
``skirt"  of thermal atoms that extends beyond the radius of the
condensate. In practice, it is difficult to extend this
measurement below about $T/T_{c} = 0.4$, except in very special
cases (for instance when a Feshbach resonance is used to set the
scattering length to zero).  For low temperatures, the density of
thermal atoms becomes so low that they are difficult to image.
Moreover, when the temperature becomes lower than or comparable to
the chemical potential of the self-interacting condensate, the
spatial extent of the thermal cloud is no longer appreciably
larger than the condensate itself.
\par
It was suggested that vortex cores might serve as ``thermal-atom
concentration pits", in order to enhance thermometry at low
temperatures.  In a simple Hartree-Fock (HF) picture of the
interaction between thermal atoms and the condensate, the
condensate density represents a repulsive interaction potential to
the thermal atoms. Along the nodal line of a vortex core, the
condensate density and presumably its repulsive interaction
potential vanish. Thus, the thermal atoms would experience the
lowest combined interaction and magnetic potential within the
cores of vortices. As a result their density would be highest
there. Additionally, images of thermal atoms in the vortex core
could be taken against a vanishing background condensate density.
Moving beyond the HF approximation, one finds a more complicated
picture. The Bogoliubov spectrum of very long wave-length thermal
phonons extends all the way down to the chemical potential. One
should contrast this energy with the energy of a thermal atom
confined to a vortex core. Perhaps the atom experiences no
interaction energy. However, the kinetic energy cost of bending
its wave function to fit inside a core with a radius of the order
of healing length must, by definition, be comparable to the
chemical potential.  In the limit of very elongated vortex cores,
there can be very low-energy, core-bending modes
\cite{DalibardKv,paultilt}. Thermal excitations of these modes
would manifest as a temperature-dependent contrast ratio. We
expect this effect is unlikely to be important in the relatively
flattened geometry of our highly rotating condensates. In any
case, without more rigorous analysis, it is not easy to predict
how the contrast ratio of our vortices should vary with
temperature, but we nonetheless set out to do a preliminary study
of the effect.
\par
We vary the final condensate temperature by changing our
rf-evaporation end point. This produces a cloud with temperatures
between $5-50~nK$ or $T/T_{c}$ between 1 and less than 0.4. Here
$T_{c}$ is calculated from the trap frequencies and a measurement
of total atom number using the formula
$T_{c}=0.94\hbar\tilde\omega_{ho}N^{1/3}$, where
$\tilde\omega_{ho}$ has been adjusted for rotation according to
the equation
$\tilde\omega_{ho}=\omega_{ho}(1-\Omega^{2}/\omega_{\rho}^{2})^{1/3}$.
When possible, $T$ is extracted from a two-component fit to the
in-trap image. Because our rotation rate and temperature are
linked through the 1D evaporative process, it is unavoidable that
$\Omega$ also varies during the data set.
\par
To measure core contrast, we expand the cloud using the usual
expansion procedure. The atom cloud is expanded radially by a
factor of 13 to ensure that the cores are large compared to our
imaging resolution. However, because we no longer care about the
precise core size we do not suppress the axial expansion.
Additionally, the axial expansion actually reduces background
fluctuations in the measured core contrast. With a factor of two
axial expansion, cores become much rounder and clearer as shown in
Fig.~\ref{Vimages}(c).  These changes allow us to achieve a higher
core contrast and quieter signal than we can without expansion.
\par
The term core brightness (1-contrast ratio) will be our metric for
this experiment. We define core brightness ($\mathcal{B}$) as
$n_{2D}(core)/n_{2D}(cloud)$, where $n_{2D}(core)$ is the observed
atom density, integrated along the line of sight, at the core
center, and $n_{2D}(cloud)$ is the projected integrated atom
density at the same point, based on a smoothed fit to the overall
atom cloud. To determine $n_{2D}(cloud)$, we fit the condensate
image to a Thomas-Fermi profile and the surrounding thermal atoms
to a Bose distribution. We find $n_{2D}(core)$ by fitting each
vortex with a Gaussian to determine its center and then averaging
five pixels around the center point to determine the integrated
density. Brightness is calculated for each vortex and then
averaged with other vortices in the cloud. To suppress noise from
low signal, vortices further than 0.4 $R_{\rho}$ from the
condensate center are disregarded for this measurement. The
$n_{2D}(core)$ term necessarily contains signal from the
surrounding thermal atoms because the vortices do not penetrate
the thermal component. Thus, one expects to see a steady decrease
in $\mathcal{B}$ with decreasing temperature, as atoms not
necessarily in the vortex core, but still in the integrated line
of sight, disappear. One would hope that $\mathcal{B}$ continues
to decrease even for $T/T_{c}$ below 0.4 for this analysis to be a
viable means of extending condensate thermometry.
\par
We are in the awkward position of comparing our core contrast
measurement to a temperature measurement that, as previously
described, is expected to fail at low temperatures. To monitor
this failure, we calculate a simplistic core brightness
($\mathcal{B}_{simple}$) found by comparing the fitted in-trap
condensate and thermal cloud profiles.  Here
$\mathcal{B}_{simple}\equiv \tilde n_{2D}(thermal)/ (\tilde
n_{2D}(condensate)+\tilde n_{2D}(thermal))$ where $\tilde
n_{2D}(condensate)$ and $\tilde n_{2D}(thermal)$ are the smoothed
condensate and thermal cloud profiles integrated along the z-axis
and averaged over a region of radius less than $0.4~R_{\rho}$ from
the condensate center. The term $\mathcal{B}_{simple}$ can be
thought of as the core brightness one would expect based on the
undoubtedly false assumption that the condensate and thermal atoms
do not interact. It is interesting to compare $\mathcal{B}$ to
$\mathcal{B}_{simple}$ since this same dubious assumption is
implicit in the standard thermometry technique of fitting the
thermal ``skirt".
\par

\begin{figure}
\begin{center}
\psfig{figure=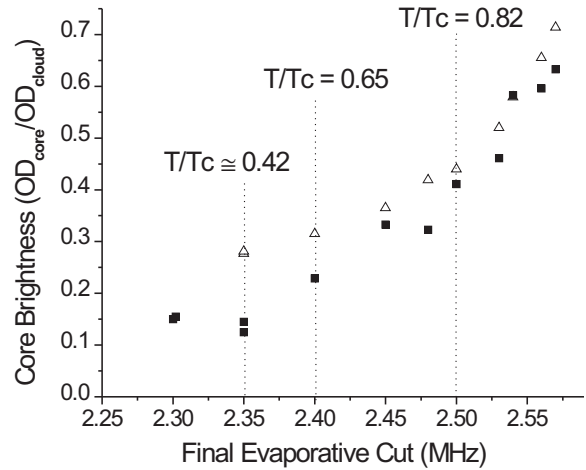,width=1\linewidth,clip=}
\end{center}
\caption {Measured core brightness as a function of final rf
evaporative cut.  Within our ability to measure, $T/T_{c}$
decreases continuously with the rf frequency. For the black
squares brightness ($\mathcal{B}$) is defined as the 2D atom
density at the vortex core divided by the 2D atom density of the
overall smoothed condensate plus thermal cloud profile at the same
point. For the open triangles a simplistic brightness
($\mathcal{B}_{simple}$) is calculated from the ratio of the 2D
atom density of the thermal cloud to the 2D atom density of the
overall smoothed condensate and thermal cloud profile. At high
temperatures $\mathcal{B}$ and $\mathcal{B}_{simple}$ exhibit a
clear dependance on the final rf cut. At lower temperatures it is
encouraging that as $\mathcal{B}_{simple}$, begins to fail
$\mathcal{B}$ is still continuing a smooth trend downward.
Disappointingly at very low temperatures, $\mathcal{B}$ plateaus
at about 0.14.} \label{Temp}
\end{figure}

In Fig.~\ref{Temp}, $\mathcal{B}$ and $\mathcal{B}_{simple}$ are
plotted versus the final evaporative cut. For our experiment, the
thermal cloud can be reliably fit for $T/T_{c} > 0.6$ and less
reliably fit for $T/T_{c} > 0.4$. In both these regions $T/T_{c}$
decreases continuously with lower final evaporative cut. It is
assumed that for $T/T_{c}$ just below 0.4, this trend continues.
For reference, three values of $T/T_{c}$ (measured from the
thermal ``skirt") are included in the plot. One can see that
$\mathcal{B}$ does steadily decrease with lower temperature for
$T/T_{c} > 0.4$. It is interesting to note that
$\mathcal{B}_{simple}$ closely tracks $\mathcal{B}$ at the higher
temperatures and then diverges from $\mathcal{B}$ as the cloud
gets colder. Presumably, this divergence occurs because thermal
atoms are pushed away from the condensate center as interactions
between the condensate and the thermal cloud become important. The
fact that $\mathcal{B}_{simple}$ diverges upwards is likely due to
the tendency of our fitting technique to overestimate the thermal
cloud density at high condensate fractions.  The failure of
$\mathcal{B}_{simple}$ at low temperatures also throws into
suspicion the quoted $T/T_{c}$ since they are determined from the
same two-component fit.
\par
In contrast, as $\mathcal{B}_{simple}$ begins to fail,
$\mathcal{B}$ continues its previous smooth downward trend. It is
also interesting to note that at an rf of 2.35~MHz, we see a
$\mathcal{B}$ of 0.13-0.15, which is not that far off from the
work of Virtanen \emph{et al.} \cite{SalomaaCore} who predict that
atoms trapped in the core would lead to a $\mathcal{B}$ of 0.1 at
a $T/T_{c}$ of $0.39$. Unfortunately, our efforts to observe a
$\mathcal{B}$ of less than 0.125 have failed so far, as can be
seen from the data points at 2.3~MHz in Fig.~\ref{Temp}. This
limit impedes our ability to measure temperatures colder than
$0.4~T/T_{c}$. Currently, it is unclear what the source of this
limit is. Perhaps the same imaging systematics that make our
vortex radius unreliable at the 10\% level are also preventing us
from seeing a core brightness level less than 0.13 or a very
slight tilt of the vortices may occur during expansion.
\par
As a caveat to the previous discussion, the same limitations that
inhibit condensate thermometry below $T/T_{c}$ of 0.4 will also
reduce the efficacy of evaporative cooling in the same regime.
Additionally, the already inefficient 1D nature of our evaporation
would exacerbate such a cooling problem. Perhaps the simplest
explanation for the failure of $\mathcal{B}$ to decrease with very
deep rf cuts is that the condensate fraction is no longer
increasing.  One could imagine the our measured $\mathcal{B}$ is
faithfully following the temperature we achieve.
\par
In summary the conclusions of our preliminary attempt to extend
thermometry with core brightness are encouraging but ambiguous.
New ideas are needed before we can make further progress.

\par
The work presented in this paper was funded by NSF and NIST. We
acknowledge fruitful discussions with Dan Sheehy and Leo
Radzihovsky. Numerical simulations were performed with the JILA
Keck cluster.
\par


\begin{thebibliography}{Hi!}

\bibitem[*]{qpdNIST}
Quantum Physics Division, National Institute of Standards and Technology.

\bibitem{DalibardLatt}
K.~W. Madison, F. Chevy, W. Wohlleben, and J. Dalibard, Phys. Rev.
Lett. {\bf 84},  806  (2000).

\bibitem{KetterleLatt}
J.~R. Abo-Shaeer, C. Raman, J.~M. Vogels, and W. Ketterle, Science
{\bf 292}, 476 (2001).

\bibitem{FootVortex}
E. Hodby, G. Hechenblaikner, S. A. Hopkins, O. M. Marago, and C.
J. Foot, Phys. Rev. Lett. {\bf 88},  010405  (2002).

\bibitem{Paul}
P. C. Haljan, I. Coddington, P. Engels, and E. A. Cornell, Phys.
Rev. Lett. {\bf 87}, 210403  (2001).

\bibitem{DalibardKv}
V. Bretin, P. Rosenbusch, F. Chevy, G. V. Shlyapnikov, and J.
Dalibard, Phys. Rev. Lett. {\bf 90}, 100403  (2003).

\bibitem{SandroKv}
F. Chevy and S. Stringari, Phys. Rev. A. {\bf 68}, 053601  (2003).

\bibitem{StoofKv}
R. A. Duine and H. T. C. Stoof, Phys. Rev. lett. {\bf 91}, 150405
(2003).

\bibitem{AnglinTk}
J.~R. Anglin and M. Crescimanno, cond-mat/0210063.

\bibitem{JilaTk}
I. Coddington, P. Engels, V. Schweikhard, and E. A. Cornell, Phys.
Rev. Lett. {\bf 91}, 100402 (2003).

\bibitem{BaymTk}
G. Baym, Phys. Rev. Lett. {\bf 91}, 110402 (2003).

\bibitem{BigelowTk}
S. Choi, L. O. Baksmaty, S. J. Woo, and N. P. Bigelow, Phys. Rev.
A. {\bf 68}, 031605 (2003).

\bibitem{MachidaTk}
T. Mizushima, Y. Kawaguchi, K. Machida, T. Ohmi, T. Isoshima, and
M. M. Salomaa, Phys. Rev. Lett. {\bf 92}, 060407 (2004).

\bibitem{BaymTk2}
S. A. Gifford, G. Baym, cond-mat/0405182.

\bibitem{Jilastripes}
P. Engels, I. Coddington, P.~C. Haljan, and E.~A. Cornell, Phys.
Rev. Lett. {\bf 89}, 100403 (2002).

\bibitem{GiantVortex}
P. Engels, I. Coddington, P. C. Haljan, V. Schweikhard, and E. A.
Cornell, Phys. Rev. Lett. {\bf 90}, 170405 (2003).

%%%%%%%%%%%%%%%%%%%%%%%%%%%%%%%%%%%%%%%%%%%%%%%%%%%%%%%%%
\bibitem{FederNv}
D. L. Feder and C. W. Clark, Phys. Rev. Lett. {\bf 87}, 190401
(2001).

\bibitem{Dan}
D. E. Sheehy and L. Radzihovsky, cond-mat/0402637.

\bibitem{BaymLatt}
G. Watanabe, G. Baym, and C. J. Pethick, cond-mat/0403470.

\bibitem{ReadLatt}
N.R. Cooper, S. Komineas, and N. Read, cond-mat/0404112.

\bibitem{FetterCores}
A. Fetter, in \emph{Lectures in Theoretical Physics}, eds. K.
Mahanthappa and W.E. Brittin, Vol. XIB, p. 351.

\bibitem{MachidaCore}
T. Isoshima and K. Machida, Phys. Rev. A. {\bf 59}, 2203 (1999).

\bibitem{SalomaaCore}
S. M. M. Virtanen, T. P. Simula, and M. M. Salomaa, Phys. Rev.
Lett. {\bf 86}, 2704 (2001).

\bibitem{HoLLL}
T. L. Ho, Phys. Rev. Lett. {\bf 87}, 060403 (2001).

\bibitem{Fischer}
U. R. Fischer and G. Baym, Phys. Rev. Lett. {\bf 90}, 140402
(2003).

\bibitem{BaymLLL}
G. Baym, C. J. Pethick, cond-mat/0308325.

\bibitem{AllanQH}
J. Sinova, C. B. Hanna, and A. H. MacDonald, Phys. Rev. Lett. {\bf
90}, 120401 (2003).

\bibitem{FetterLLL}
A.L. Fetter, Phys. Rev. A. {\bf 64}, 063608 (2001).

\bibitem{SandroTc}
S. Stringari, Phys. Rev. Lett. {\bf 82}, 4371 (1999).

%%%%%%%%%%%%%%%%%%%%%%%%%%%%%

\bibitem{JilaLLL}
V. Schweikhard, I. Coddington, P. Engels, V. P. Mogendorff, and E.
A. Cornell, Phys. Rev. Lett. {\bf 92}, 040404 (2004).

\bibitem{SandroSc}
D. Guery-Odelin and S. Stringari, Phys. Rev. Lett. {\bf 83}, 4452
(1999).

\bibitem{FootSc}
O. M. Marago, S. A. Hopkins, J. Arlt, E. Hodby, G. Hechenblaikner,
and C. J. Foot, Phys. Rev. Lett. {\bf 84}, 2056 (2000).

\bibitem{EllipEqn}
Ellipticity defined as
$(\omega_{+}^{2}-\omega_{-}^{2})/(\omega_{+}^{2}+\omega_{-}^{2})$
where $\omega_{+}$ and $\omega_{-}$ are the trap frequencies in
the along the major and minor trap axes respectively.

\bibitem{Heather}
H. J. Lewandowski, D. M. Harber, D. L. Whitaker, E. A. Cornell, J.
Low Temp. Phys. {\bf 132}, 309 (2003).

\bibitem{Modugno}
F. Dalfovo, M. Modugno, Phys. Rev. A. {\bf 61}, 023605 (2000).

\bibitem{Castin}
Y. Castin and R. Dum , Phys. Rev. Lett. {\bf 77}, 5315 (1996).

\bibitem{2d}
The radial expansion decreases the density, and thus is likely to
increase the accuracy of the mean-field treatment. However, once
the density has decreased past the point where the chemical
potential is small compared to the kinetic energy associated with
the finite axial extent the gas enters an effectively 2D regime.
In this regime the extent to which microscopic atom-atom
correlations subvert the accuracy of the mean-field approximation
becomes indendent of further reductions in density.  This is in
contrast to the 3D regime, in which lower density always means
better mean-field accuracy, and also in contrast to the 1D regime,
in which lower density actually promotes mean-field failure. For
the parameters of our experiment (in particular for the case that
$a_{sc} \ll R_z$), the effects of correlations in the 2D regime
are very small. We thank an anonymous referee for prompting us to
look more closely at this limit.

\bibitem{allan}
A. H. MacDonald, private communication.

\bibitem{AspectTc}
F. Gerbier, J. H. Thywissen, S. Richard, M. Hugbart, P. Bouyer,
and A. Aspect, Phys. Rev. Lett. {\bf 92}, 030405 (2004).

\bibitem{paultilt}
P. C. Haljan, B. P. Anderson, I. Coddington, and E. A. Cornell,
Phys. Rev. Lett. {\bf 86}, 2922 (2001).

\end{thebibliography}
\end{document}